# 3D-Aspekte des Materialdesigns metallischer Werkstoffe

## *3D Aspects of Materials Design for Metals and Alloys*

U. Krupp, C. Haase, W. Song, W. Bleck, Aachen[1]
K. Jahns, Osnabrück[2]

**Abstract**

Die Bemessung technischer Strukturbauteile beruht auch heute noch in vielen Fällen auf der Annahme isotroper Materialeigenschaften, die in kontinuumsmechanische Modelle einfließen. Demgegenüber stehen komplex aufgebaute Werkstoffe, deren Mikrostruktur durch moderne Ur- und Umformverfahren eine ausgeprägte Ortsabhängigkeit in verschiedenen Längenskalen aufweist. Additiv gefertigte Strukturen ermöglichen eine bionisch inspirierte, lastangepasste Verteilung des Materials. Durch gezielte Temperaturführung bei der Erstarrung oder der Wärmebehandlung können Anisotropie oder Ausscheidungseffekte bis in den atomaren Maßstab zur Werkstoffoptimierung genutzt werden. Der Beitrag zeigt anhand von Beispielen die Bedeutung und Umsetzung der dreidimensionalen Korrelation von Mikrostruktur und Werkstoffeigenschaften.

*Nowadays, most structural integrity concepts rely on simplified isotropic material data that are used within continuum mechanics modeling approaches. In contrast, modern casting and forming processes yield complex microstructures coming along with pronounced gradients in the material's properties in various length scales. By means of additive manufacturing. bionics-inspired structures can be perfectly adapted to the space-resolved load distribution. Specific temperature control during solidification or heat treatment allow for usage of anisotropy and precipitation effects for materials optimization down to the atomic length scale. The present paper shows the significance and practical application of a three-dimensional correlation between the material's microstructure and its technical properties.*

## 1　Einleitung

Der Aufbau technischer Werkstoffe ist durch eine komplexe Dreidimensionalität in mehreren Längenskalen geprägt. Kristallite bzw. Körner unterschiedlicher Größe, Form, Struktur und kristallographischer Orientierung bilden das Gefüge. Bedingt durch die Kristallstruktur sind die Werkstoffeigenschaften richtungsabhängig (anisotrop). Eine mechanische Beanspruchung zwingt die einzelnen Körner dazu, sich kompatibel zu verformen, was zu einer räumlich inhomogenen Spannungsverteilung führt. Dabei spielen Korn- oder Phasengrenzen eine besondere Rolle. Durch Defektanreicherungen (Versetzungen, Leerstellen, Segregation von Fremdatomen, Ausscheidungen) handelt es sich bei ihnen einerseits um Schwachstellen und/oder schnelle Diffusionspfade, andererseits bilden sie Barrieren gegenüber Versetzungsbewegung, die aus der Überschreitung der kritischen Schubspannung durch die lokal anliegende Schubspannung resultiert. Diese Effekte lassen sich für die gezielte Einstellung anisotroper Eigenschaften nutzen, bspw. für die Drahtherstellung, für gerichtet erstarrte Turbinenschaufeln oder das sogenannte Korngrenzen-Engineering zur Entwicklung bspw. spannungsrisskorrosionsbeständiger Werkstoffe [1]. Bei letztgenanntem ist die dreidimensionale Perkolation von speziellen Koinzidenz-Korngrenzen zur Unterbrechung eines durchlässigen Netzwerks schwacher, zufällig misorientierter Korngrenzen von Bedeutung [2].

[1] Prof. Dr.-Ing. habil. Ulrich Krupp, Dr.-Ing. Christian Haase, Dr.-Ing. Wenwen Song, Prof. Dr.-Ing. Wolfgang Bleck, Institut für Eisenhüttenkunde der RWTH Aachen.
[2] Dr.-Ing. Katrin Jahns, Laborbereich Materialdesign und Werkstoffzuverlässigkeit, Hochschule Osnabrück.

Innerhalb der Kristallite können dreidimensional ausgerichtete Zweit- oder Drittphasen oder atomare Cluster-Anordnungen das Werkstoffverhalten maßgeblich bestimmen. Eine Erhöhung des Widerstands gegenüber Versetzungsbewegung erhöht die Werkstofffestigkeit, führt aber gleichzeitig zu einer Reduktion der Duktilität. Demgegenüber stehen metastabile Phasen, die infolge einer Verformung umwandeln und so eine in-situ Festigkeitssteigerung an hoch belasteten Stellen bewirken (bspw. verformungsinduzierte martensitische Umwandlung in TRIP-Stählen). Austenitische Phasen mit niedriger Stapelfehlerenergie tendieren bei Belastung zur Zwillingsbildung. Die dabei neu entstehenden Grenzflächen bilden zusätzliche Barrieren für die weitere Versetzungsbewegung (TWIP-Effekt in Hochmanganstählen [3]).

Der Beitrag zeigt an drei Beispielen, welche Rolle und welches Nutzungspotential die dreidimensionale Ausprägung der Makro-, Meso- und Mikrostruktur für die Schädigungsmodellierung, die Nanostrukturierung und die additive Fertigung metallischer Konstruktionswerkstoffe als Forschungsschwerpunkte am Institut für Eisenhüttenkunde (IEHK) spielt.

## 2   3D-Aspekte der Ermüdungsschädigung und Modellanforderungen

Zyklische Beanspruchung mit Lastamplituden unterhalb der makroskopischen Streckgrenze führt aufgrund der elastischen Ansiotropie zu lokalen Spannungsüberhöhungen und Versetzungsgleiten. Die zyklische Plastizität weist einen irreversiblen Anteil auf, der entlang von Gleitbändern zu Defekten und darauf folgend zu Extrusionen und Intrusionen akkumuliert (siehe Bild 1a, martensitischer Stahl 50CrMo4). Mit fortschreitender Zyklenzahl kommt es an diesen zur Rissbildung (Bild 1b). Im Gegensatz zu den Annahmen der linear elastischen Bruchmechanik breiten sich diese Risse nicht halbkreisförmig, sondern in einem ausgeprägten dreidimensionalen Zusammenhang mit der lokalen Mikrostruktur aus. Die Schädigung orientiert sich an der räumlichen Anordnung der Körner und Phasen. Das Beispiel in Bild 1 zeigt, dass sich die Risse an der Oberfläche und in der Tiefe an den {110}-Ebenen des angelassenen Martensits orientieren.

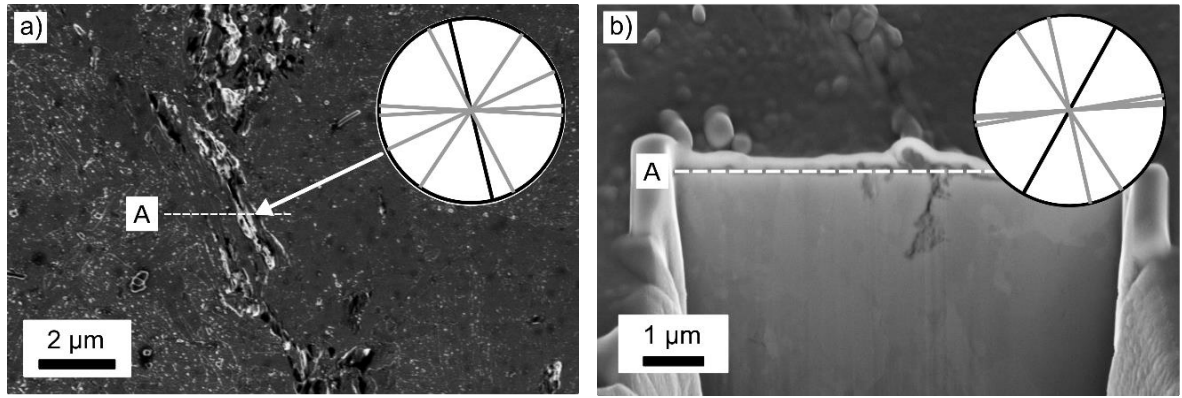

**Bild 1:** Extrusionen und Intrusionen entlang von {110}-Gleitebenen im Stahl 50CrMo4 ($\sigma_a$=600MPa, $R$=-1, die mittels EBSD abgeleiteten Gleitspuren sind in den Kreisen markiert): (a) Oberfläche, (b) FIB-Schnitt entlang der Linie A.

Die für die dreidimensionale Schädigungsanalyse erforderlichen räumlichen Abbildungstechniken müssen über eine hinreichende Auflösung und die Möglichkeit zur Separation von Phasen und einzelnen Körnern verfügen. Alternative Techniken sind (i) serielles Schneiden - entweder metallographisch oder durch den Ionenstrahl im FIB-REM (focussed ion beam-Rasterelektronenmikroskop, s. Bild 1b) - in Kombination mit EBSD (electron back-scatter diffraction) oder (ii) die Röntgen/Synchrotron-Tomographie. Bild 2 zeigt anhand einer wechselverformten Probe des Duplexstahls 1.4462 (vgl. [4]) die mikrostrukturkontrollierte Ermüdungsrissausbreitung entlang der Oberfläche (Bild 2a) und die mittels Synchrotron-Computertomographie (CT) dargestellte Rissentwicklung im Materialinnern (Bild 2b). Wechselwirkungen mit der lokalen Mikrostruktur lassen sich dabei durch Anwendung der Phasenkontrast-Tomographie (PCT) und der Diffraktionskontrast-Tomographie (DCT) nachweisen. Während der Rotation der Probe zur CT-Darstellung werden von Kristalliten, die die Bragg-Bedingung erfüllen, Röntgenintensitäten abgebeugt. Durch Auswertung der Abbildungsorte der abgebeugten Intensitäten mit den im jeweiligen CT-Schnitt dunkler erscheinenden Bereichen kann die dreidimensionale kristallographische Orientierungsverteilung dargestellt werden (vgl. [5]).

Mit Hilfe der dreidimensionalen Mikrostrukturdaten können (i) die aus der elastischen und plastischen Anisotropie resultierende inhomogene Spannungsverteilung (mittels kristallplastischer Finite Elemente-Methode, CP FEM) und (ii) die Entwicklung der akkumulierten zyklischen Plastizität in Verbindung mit Rissbildung und Ermüdungsrissausbreitung berechnet werden (vgl. [2,6]). Bei letzterer bestimmt die dreidimensionale Ausrichtung der Gleitbänder in Bezug zur Korngrenzenebene (charakterisiert durch Kippwinkel Φ und Verdrehwinkel ξ) die Widerstandswirkung von Korn- und Phasengrenzen (vgl. Bild 3). Der Versetzungsaufstau zwischen Korngrenze und Rissspitze bremst die Rissausbreitung infolge einer Rückspannung ab. Sobald der in Bild 3 markierte Anteil der Korngrenzenfläche (Barriere) überwunden ist, kann sich die plastische Verformung in das Nachbarkorn B ausbreiten; die Rückspannung fällt demzufolge ab und die Rissausbreitungsrate nimmt wieder zu. Dies erlaubt eine ortsaufgelöste Berechnung der Rissausbreitungsrate *da/dN*, die auf Null absinken kann. Dies kann als Bedingung für lokale Dauerfestigkeit angesehen werden.

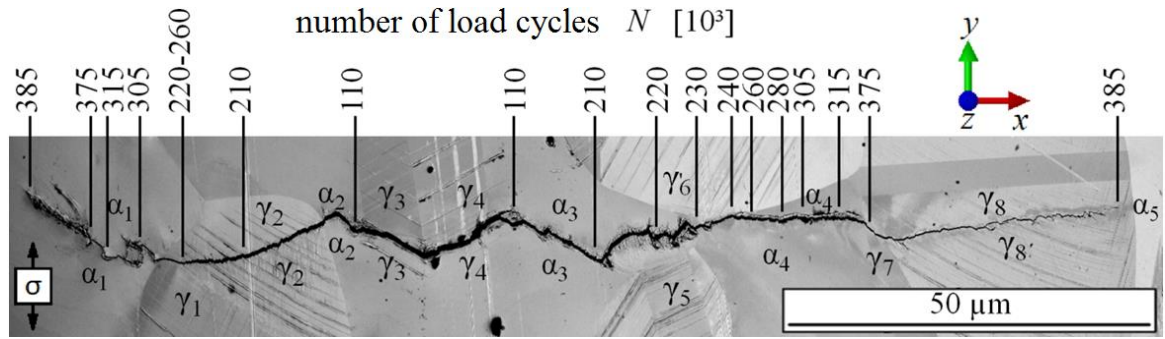

a

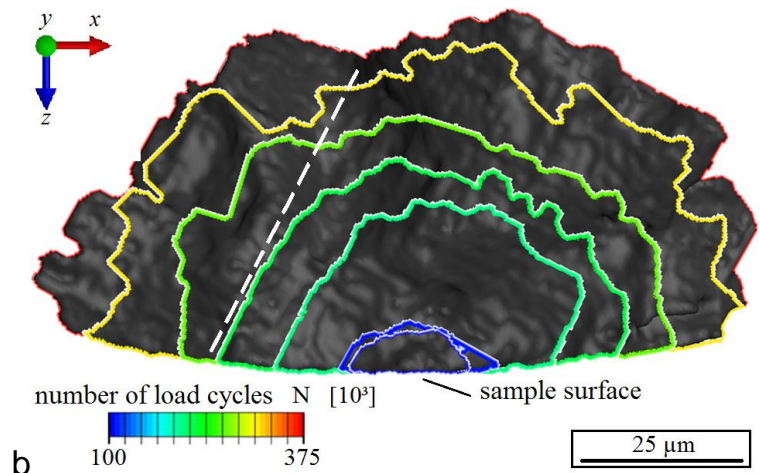

b

**Bild 2:** Verfolgung der Ermüdungsrissausbreitung in einem austenitisch ($\gamma$) - ferritischen ($\alpha$) Duplexstahl als Funktion der Zyklenzahl (*R*=-1): (a) Oberflächenrissausbreitung im REM, (b) Verlauf der Rissfront in die Tiefe (Synchrotron CT, ESRF Grenoble, W. Ludwig).

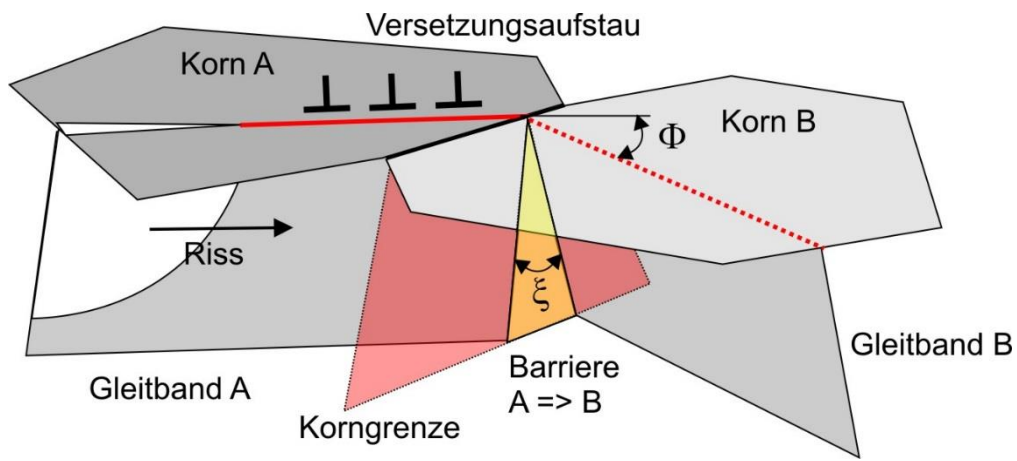

**Bild 3:** Schematische Darstellung der dreidimensionalen Wechselwirkung zwischen Ermüdungsrissausbreitung und Korngrenzen.

## 3   Materialdesign auf der Nanometer-Skala: Atomsondentomographie

Atomsondentomographie (engl. *atom probe tomography*, APT) basiert auf den Prinzipien der Feldionisation und -verdampfung zur Erzeugung einer dreidimensionalen Rekonstruktion der Elemente mit atomarer Auflösung. Die Probenpräparation in nadelförmige Spitzen mit einem Radius von weniger als 100 nm erfolgt entweder mittels eines zweistufigen Elektropoliervorgangs oder mittels ortspezifischer Extraktion mit Hilfe eines FIB-Mikroskops. Im Ultrahochvakuum wird an die Proben eine Gleichspannung von 5-20 kV [7] angelegt, um oberflächennahe Atome durch das hohe elektrische Feld zu ionisieren. Präzise Energiepulse mittels Laser oder Wechselstroms erzeugen ausreichend Energie, um die Oberflächenbindungen einzelner Atome zu überschreiten und die Verdampfung zu initiieren. Ein „*time-of-flight*" Massenspektrometer in Kombination mit einem Flächendetektor ermöglicht die räumlich differenzierte Identifikation von Elementen mit Hilfe derer Massen-zu-Ladungs-Verhältnis. Eine 3D-Rekonstruktion wird auf Basis der detektierten Ionensequenz, der Positionsdaten auf dem Flächendetektor und des Massen-zu-Ladungs-Verhältnis erzeugt. APT ermöglicht die dreidimensionale Untersuchung atomarer Zusammensetzungen von Ausscheidungen, der atomaren Clusterbildung und chemischer Gradienten an Werkstoffdefekten, beispielsweise Versetzungen oder Korngrenzen.

*Nano-skalierte Karbidbildung durch dehnungsinduzierten Martensitzerfall in Walzlagerstählen unter Rollkontaktermüdung*

Die Ausbildung weiß und dunkel ätzender Bereiche (white etching areas, WEA, dark-etching regions, DER) während Rollkontaktermüdung steht wahrscheinlich im Zusammenhang mit einer dehnungsinduzierten Kohlenstoffmigration **Fehler! Verweisquelle konnte nicht gefunden werden.**. Somit würde die Umverteilung von Kohlenstoff in der Matrix den Zerfall des Martensits kontrollieren. Zur weiteren Untersuchung der Umverteilung des Kohlenstoffs im Rahmen der Ausbildung von DER, wurden APT-Messungen am DER-Ferrit durchgeführt. Eine 3D-Atomverteilung innerhalb der DER ist in Bild 4a dargestellt. Die inhomogene Verteilung des Kohlenstoffs manifestiert sich durch kohlenstoffangereicherte und -dezimierte Bereiche. Folglich repräsentieren die kohlenstoffdezimierten Bereiche den DER-Ferrit, während die kohlenstoffangereicherten Bereiche nano-skalierte Ausscheidungen, z.B. θ-, η- und ε-Karbide, darstellen. Darüber hinaus ist Kohlenstoff an den Grenzflächen der Martensitplatten mit 6-7 at% angereichert (siehe Pfeile und gestrichelte Linie in Bild 4a). Dieser Kohlenstoffgehalt stimmt mit dem Sättigungsgehalt von Kohlenstoff in Cottrell'schen Atmosphären überein [9, 10]. Bild 4a zeigt exemplarische Elementverteilungen an den Grenzflächen von verschiedenen Ausscheidungen auf. In ausgewählten Bereichen (ROI) wurden die Konzentrationsverläufe bestimmt und in Bild 4b dargestellt. Die Ausscheidungen wurden nach dem Kohlenstoffgehalt mit ca. 25 at% als $Fe_3C$-Karbide (θ), mit ca. 33.3 at% als $Fe_2C$-Karbide (η) und mit ca. 29.4 at% als $Fe_{2.4}C$-Karbide (ε) basierend auf der jeweiligen Stöchiometrie identifiziert. Insgesamt wurden über 15 kohlenstoffangereicherte Bereiche in dem Datensatz erfasst, wobei ein Großteil als θ-Karbide identifiziert wurde. Dennoch ist die Anwesenheit von Übergangskarbiden umgeben von DER-Ferrit ein starker Hinweis auf die Kohlenstoffmigration von der jeweiligen Martensitmatrix hin zu vorhandenen ε- und η-Karbiden, die während des Vergütens gebildet wurden [11, 12].

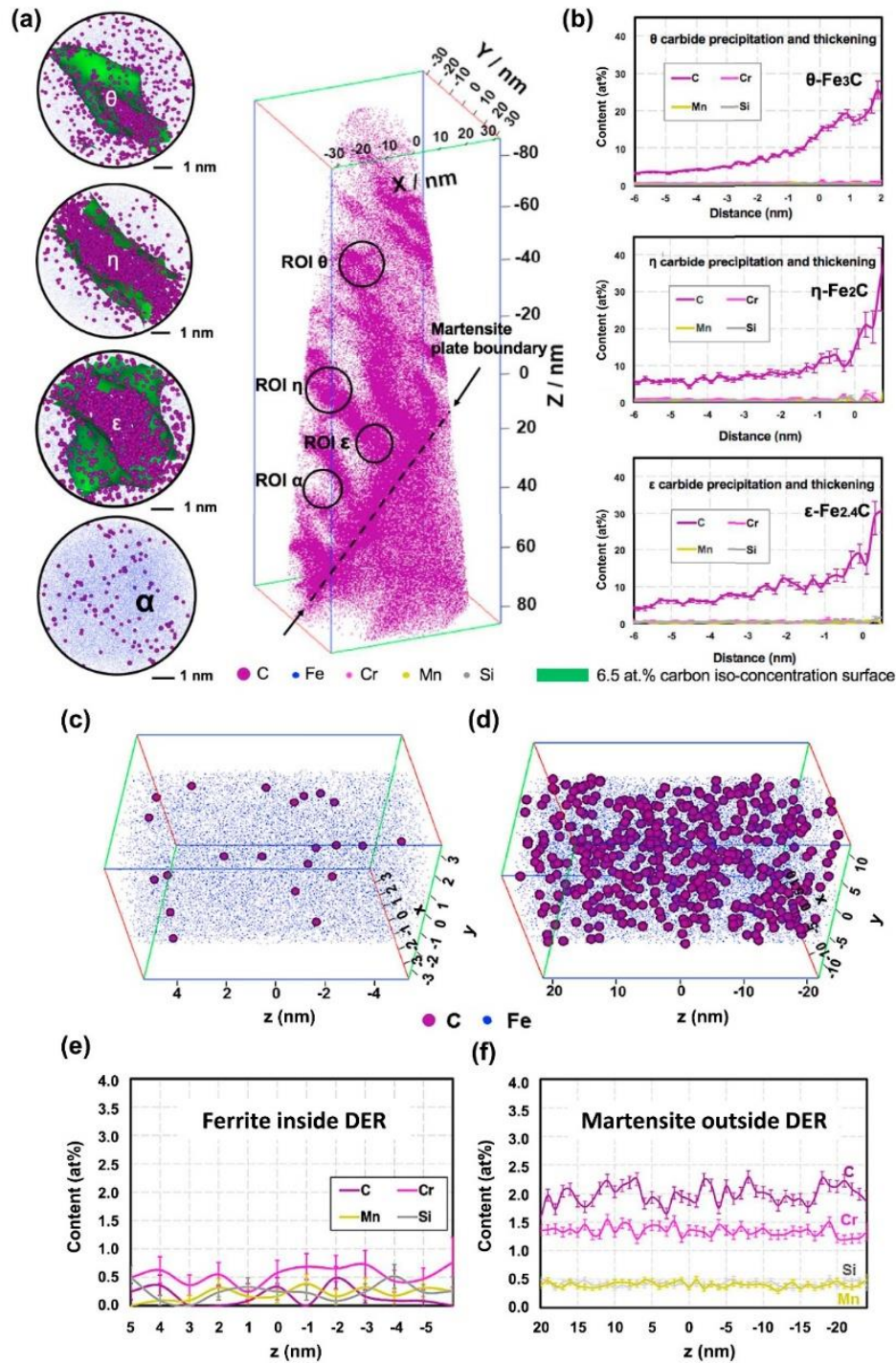

**Bild 4:** Darstellung des APT-Datensatzes als (a) 3D atomare Verteilung mit Analyse der verschiedenen Karbidtypen (ε, η, θ) und des DER-Ferrits und ausgewählter Bereiche (ROI). (b) Elementverteilung für ausgewählte Karbide in (a). (c) und (d) Vergleich der Kohlenstoffverteilung im DER-Ferrit bzw. Martensit. (e) und (f) zugehörige 1D-Konzentrationsverläufe der verschiedenen Legierungselemente.

Diese Beobachtung stützt den grundlegenden Mechanismus, der für die DER-Ausbildung postuliert wurde [8]. Zur Beurteilung der Kohlenstoffreduktion bzw. -anreicherung im DER-Ferrit bzw. im zugehörigen Martensit sind die entsprechenden Kohlenstoffverteilungen in Bild 4c bzw. Bild 4d dargestellt. Die zugehörigen 1D-Konzentrationsverläufe sind in Bild 4e bzw. Bild 4f dargestellt. Zur verbesserten Darstellung sind Kohlenstoffatome als Kugeln bzw. Eisenatome als Punkte dargestellt. Der mittlere Kohlenstoffgehalt im Martensit beläuft sich auf ca. 2.1 at%, während der Kohlenstoffgehalt im DER-Ferrit deutlich geringer ausfällt mit Gehalten von 0.02-0.06 at%. Darüber hinaus ist anzumerken, dass der Chromgehalt im DER-Ferrit (~0.66 at%) ebenfalls niedriger ausfällt im Vergleich zum Martensit (~1.39 at%). Dies könnte auf die Umverteilung von Chrom in die Umgebung von chromreichen, globularen Karbiden zurückzuführen sein.

## 4    Materialdesign durch additive Fertigung

Die Technologien der additiven Fertigung (umgangssprachlich auch als 3D-Druck-Verfahren bezeichnet) weisen zahlreiche technologische und ökologische Vorteile gegenüber umformenden und subtraktiven Methoden auf. Hierbei werden dreidimensionale Objekte aus digitalen 3D-Modellen über einen mehrfachen schichtweisen Materialauftrag aus formlosem Rohstoff hergestellt. Der schichtweise Aufbau erfolgt durch lokales Aufschmelzen des Einsatzmaterials, üblicherweise Metallpulver oder -draht, unter Zuhilfenahme einer Wärmequelle mit hoher Energie, beispielsweise Laserstrahlquellen. Da nicht nur das aufzutragende Material, sondern auch die darunterliegende Schicht partiell erneut erschmolzen wird, erfolgt eine Anbindung durch das Verschweißen benachbarter Schichten. Für das Material- und Bauteildesign ergeben sich durch Ausnutzung der folgenden prozessspezifischen Vorteile neue Freiheitsgrade [13]:

- Designfreiheit: Herstellung geometrisch komplexer Bauteile mit lastangepasster Geometrie.
- Mischung unterschiedlicher Pulverwerkstoffe: flexible Variation der chemischen Zusammensetzung.
- Flexible Steuerung der Prozessparameter: lokale Variation der Abkühlbedingungen und Mikrostruktur.

Wie in Bild 5 gezeigt ist, können mittels der additiven Fertigungsmethode des selektiven Laserstrahlschmelzens geometrisch komplexe Proben mit filigranen Strukturen gefertigt werden. Beide Proben wurden aus dem Stahl 1.4404 hergestellt, weisen eine relative Dichte im Vergleich zum Vollmaterial von 33% und Strebendicken/Wandstärken von wenigen 100 µm auf. Je nach Anwendungsfall können diese für den Leichtbau geeigneten Strukturen bezüglich ihrer Verformungsmodi gezielt eingestellt werden. Aufgrund der senkrechten Streben der $f_{2cc,z}$-Gitterstruktur (1) verformt sich diese im Druckversuch durch einen dehnungsdominierten Verformungsmodus und eignet sich somit für Bauteile mit energieabsorbierender Funktion. Im Gegensatz dazu zeigt die Probe mit Hohlkugelstruktur (2) ein biegungsdominiertes Verhalten mit guten Dämpfungseigenschaften [14].

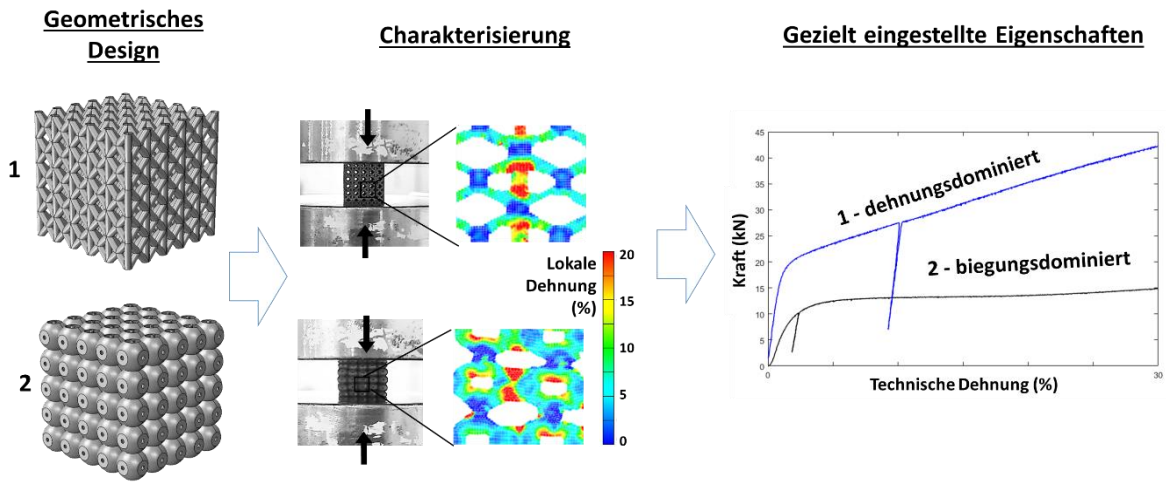

**Bild 5:** Beeinflussung des Verformungsverhaltens durch dreidimensionales geometrisches Probendesign. Die $f_{2cc,z}$-Gitterstruktur (1) weist einen Verformungsmodus auf, der durch plastische Dehnung charakterisiert ist, wohingegen die Probe mit Hohlkugelstruktur (2) biegungsdominiert verformt.

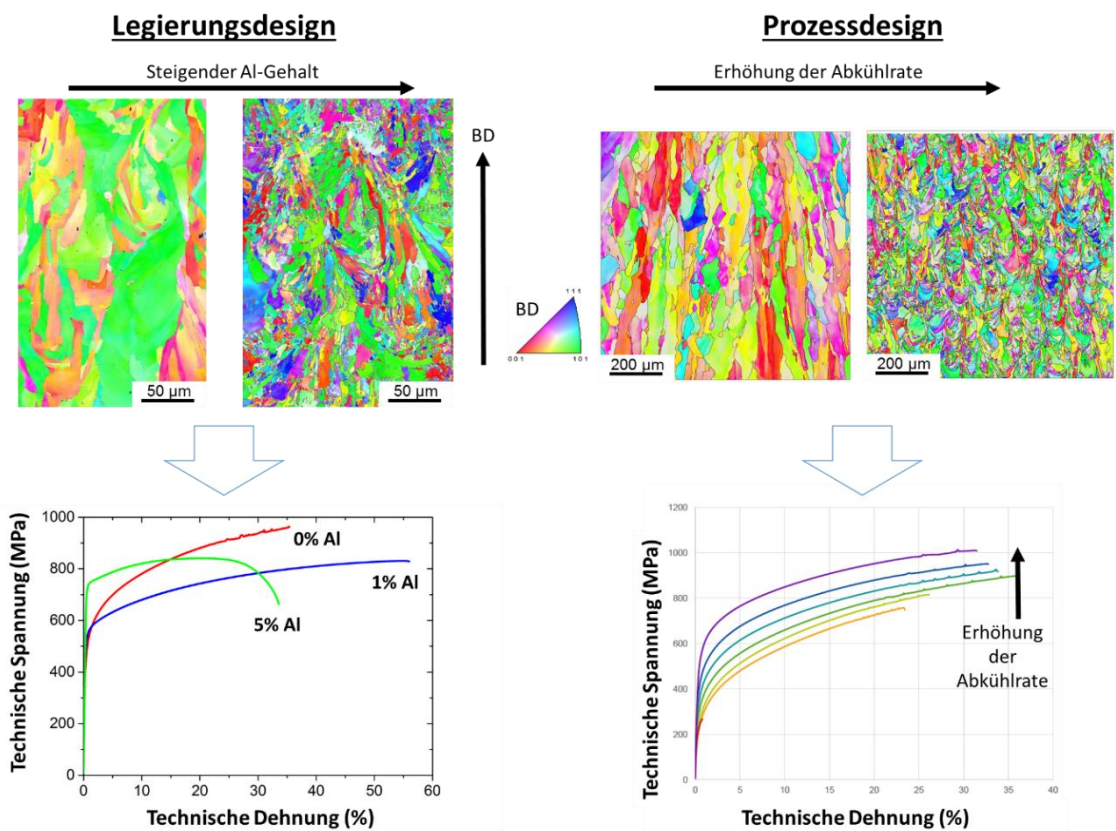

**Bild 6:** Möglichkeiten zur Steuerung der mechanischen Eigenschaften über Legierungs- und Prozessdesign. Sowohl der Al-Gehalt (linke Seite) als auch die Abkühlbedingungen (rechte Seite) können effektiv genutzt werden, um die Mikrostruktur und mechanischen Eigenschaften zu kontrollieren (Werkstoff: Stahl X30Mn22).

Das flexible Mischen unterschiedlicher Pulver sowie die einfache Anpassung der Prozessparameter erlauben weiterhin, die Mikrostruktur von additiv gefertigten Werkstoffen zu steuern. Am Institut für Eisenhüttenkunde werden diese Möglichkeiten unter anderem für unterschiedliche Stähle, Hochentropielegierungen und Kupferbasiswerkstoffe (Technologie-Campus 3D-Materialdesign Osnabrück), untersucht. Bild 6 verdeutlicht hierzu die Möglichkeiten durch gezieltes Legierungs- und Prozessdesign am Beispiel eines Hochmanganstahls X30Mn22. Durch Erhöhung des Al-Gehalts kann neben der Stapelfehlerenergie auch die Kornstruktur und somit die mechanischen Eigenschaften stark beeinflusst werden (Bild 6, links). Einerseits bewirkt die Zugabe von Al die Erhöhung der Stapelfehlerenergie und somit eine Verringerung der Verfestigungsrate. Andererseits kann durch Stabilisierung der Ferritphase eine deutliche Kornfeinung durch Vermeidung von epitaxialem Kornwachstum erreicht werden, was die Streckgrenze erhöht. Zusätzlich kann die Mikrostruktur durch Wahl geeigneter Prozessparameter kontrolliert werden (Bild 6, rechts). Hierbei ist ein deutlicher Übergang von groben kolumnaren Körnern mit starker Vorzugsorientierung zu feineren gleichachsigen Körnern mit regelloser Orientierungsverteilung erkennbar. Dies bewirkt eine deutliche Erhöhung der Festigkeit des untersuchten Stahls [15].

## 5  Schlussbetrachtung

Moderne Fertigungsverfahren, wie das selektive Laserstrahlschmelzen oder innovative Umformtechnologien, erlauben die gezielte dreidimensionale Anpassung der Werkstoffmikrostruktur an die Beanspruchung von bzw. Anforderungen an komplexe Bauteile. Ziel der Arbeiten am Institut für Eisenhüttenkunde ist die skalenübergreifende experimentelle und modellmäßige Beschreibung der dreidimensionalen Zusammenhänge zwischen Fertigungsprozessen, Mikrostruktur und Eigenschaften für ein effizientes Materialdesign.

## 6  Literatur